\documentclass[
    ,final            % use final for the camera ready runs
  ]
  {aipproc}

\layoutstyle{8x11double}

\begin{document}

\title{Monte-Carlo studies of the angular resolution of a future
  Cherenkov gamma-ray telescope}


\author{S. Funk}{
  address={Kavli Institute for Particle Astrophysics and Cosmology,
    Stanford, CA-94025, USA}
}

\author{J.A. Hinton}{
  address={School of Physics \& Astronomy, University of Leeds, Leeds,
    LS2 9JT, UK}
}

\begin{abstract}
  The current generation of Imaging Atmospheric telescopes (IACTs) has
  demonstrated the power of this observational technique, providing
  high sensitivity and an angular resolution of $\sim$0.1$^{\circ}$
  per event above an energy threshold of $\sim$100~GeV. Planned future
  arrays of IACTs such as AGIS or CTA are aiming at significantly
  improving the angular resolution. Preliminary results have shown
  that values down to $\sim 1'$ might be achievable. Here we present
  the results of Monte-Carlo simulations that aim to exploring the
  limits of angular resolution for next generation IACTs and
  investigate how the resolution can be optimised by changes to array
  and telescope parameters such as the number of pixel in the camera,
  the field of view of the camera, the angular pixel size, the mirror
  size, and also the telescope separation.
\end{abstract}

\maketitle

%%%%%%%%%%%%%%%%%%%%%%%%%%%%%%%%%%%%%%%%%%%%
%% MAINMATTER
%%%%%%%%%%%%%%%%%%%%%%%%%%%%%%%%%%%%%%%%%%%%

\section{The approach for studying the angular resolution}

\begin{figure}
  \includegraphics[height=.3\textheight]{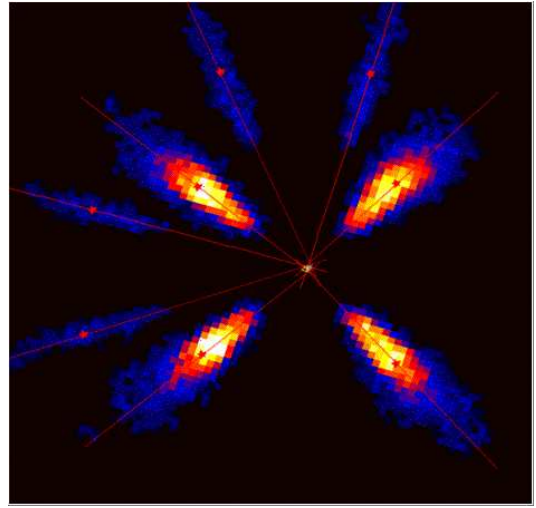}
  \caption{Reconstruction algorithms, taking the intersection of the
    major axes (Hillas analysis). In this case 8 telescopes
    participated in the event.}\label{fig::Fig1}
\end{figure}

Monte-Carlo simulations of the development of gamma-ray-induced
showers in the atmosphere were performed using Corsika
v6.2041~\citep{Corsika}. Cherenkov photons produced in the shower
development were recorded and in a second (post-processing) step,
these photons were collected with an array of telescopes with
adjustable parameters as described below. Using only those photons
that hit one of the telescopes in the array, the direction was
reconstructed from camera images using a Hillas-style analysis (see
Fig.~\ref{fig::Fig1}). For comparison we also applied more
sophisticated reconstruction methods such as a simultaneous
minimisation of all the image axes and shower cores.

Since the post-processing step is rather fast in comparison to the
generation of the showers in the atmosphere, the phase space of
different telescope configurations can be explored rather
quickly. Adjustable parameters in the post-processing step are:
\begin{itemize}
\item Number of telescopes in the array
\item Diameter of the telescope (mirror size)
\item Distance between telescopes
\item Field of view (FoV) of the camera
\item Angular size of the pixels
\item Light-collection efficiency of the pixels
\end{itemize}

\begin{figure}[ht]
  \includegraphics[width = 0.62\textwidth]{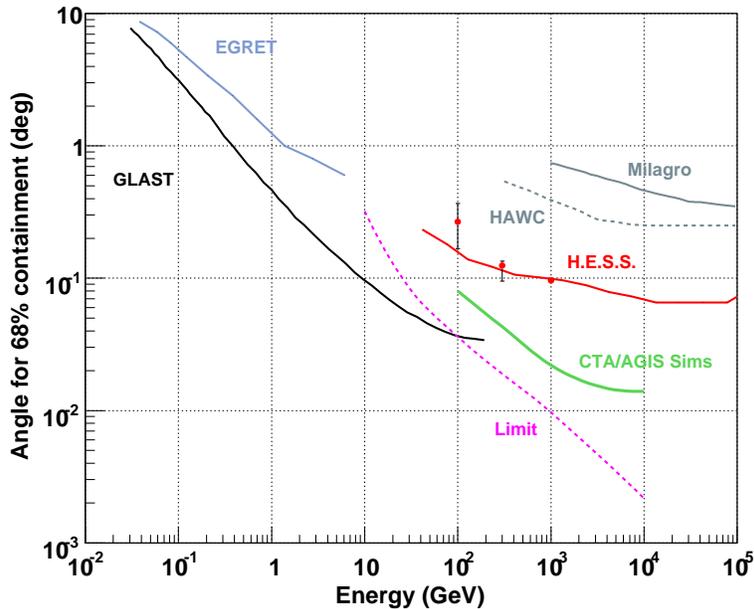}
  \caption{Angular resolution as a function of energy for different
    instruments. The H.E.S.S. curve can be reproduced in this approach
    as shown by the red circles (the red H.E.S.S.\ curve is taken from
    ~\citep{Funk:Thesis}), the dashed curve labelled 'limit' is taken
    from~\citep{Hofmann:Performance}. The green curve (CTA/AGIS) is
    derived in this study. For this simulation we used a hypothetical
    system of 49 telescopes with 30m diameter each.} \label{fig::Fig2}
\end{figure}

To verify the approach taken in this study, the first
proof-of-principle was to reproduce the angular resolution of
H.E.S.S.\ as shown in Figure~\ref{fig::Fig2} (red points). The next
step was to simulate a so-called {\sl{reference array}} with 49 30-m
telescopes with 50 m spacing, a large field of view (10$^{\circ}$) and
a pixel size of 0.06$^{\circ}$ as a test of the most optimistic
version of a future array that all other configurations could be
measured against. This reference array can also be compared to the
(optimal) case in which all Cherenkov photons emitted in the showers
are collected and used for the reconstruction as reported in
~\citep{Hofmann:Performance}. 

As can be seen from Figure~\ref{fig::Fig2} our approach gets within
$\sim$ a factor of 2--3 of that optimal angular resolution with a
value of $\sim 0.025^{\circ}$ at 1 TeV. This gives us confidence in
the approach since this reference array - albeit probably
prohibitively expensive - is by no means optimised. Apart from the
array and camera parameters that could be improved, similar to the
original study~\citep{Hofmann:Performance}, the analysis is not
optimised and could certainly be improved (in particular the
tail-cuts).

Since this paper does not allow us to address the full phase space of
possible changes to the array and camera configurations, in the
following we will focus on two of the most important variables that
affect the angular resolution of a future Cherenkov system: telescope
multiplicity (that is the number of telescopes participating in the
event) and angular pixel size. We also investigated the effect of
other parameters such as mirror size, distance between the telescopes
and light-collection efficiency of the individual photodectors but
found rather modest dependencies of the angular resolution on these
parameters (they mostly affect the energy threshold and effective
area).

\section{Angular resolution as a function of Telescope multiplicity}

\begin{figure}[hb]
  \includegraphics[width = 0.48\textwidth]{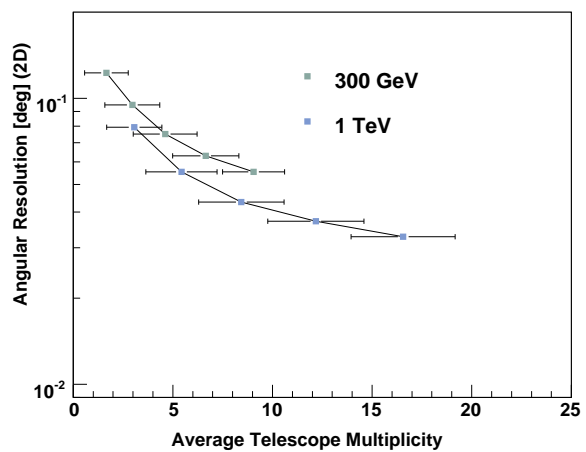}
  \caption{Angular resolution as a function of the average number of
    telescopes in the reconstruction. Plotted on the x-axis is the
    average number of telescopes that participated in the
    reconstruction. E.g. for the 49 telescope array at 1 TeV, the
    average multiplicity was ~16.} \label{fig::Fig3}
\end{figure}

Given that the directional reconstruction is performed via the
intersection of image axes, an important question that can be
addressed in our scheme is the dependence of the angular resolution on
the number of telescopes participating in the event. For this,
telescopes were randomly switched off from the reference array (49
telescopes with 50m telescope spacing) and the resulting angular
resolution was plotted as a function of the average multiplicity for
two different energies (300 GeV and 1 TeV) as shown in Fig.~\ref{fig::Fig3}. 

The 300 GeV curve does not continue to higher
multiplicities, because even for the system of 49 telescopes, the
average multiplicity for a 300 GeV shower is only ~9 (for the
reference system in which the telescopes are 50m apart). As can be
seen from these curves, the angular resolution improves rather
strongly when adding telescopes to an array of a few telescopes, but
at a large number of telescopes, it levels off as expected since the
reconstruction becomes over-constrained. This asymptotic behaviour of
the angular resolution with increasing telescope multiplicity suggests
that beyond ~10 telescopes participating in the event, the angular
resolution improves only very moderately for 1~TeV $\gamma$-rays.

\section{Angular resolution as a function of angular pixel size}

The next property of a future TeV gamma-ray array that was studied is
the dependence of the angular resolution on the (angular) size of the
individual photo-sensors.  A priori it is not obvious, whether a finer
pixelation of the camera does improve the angular resolution, since
the reconstruction is done by an intersection of image axis and
individual pixels only contribute to a reconstruction of the major
shower axis in case of a Hillas-type analysis. It is expected that
once the pixel size gets larger than the angular width of the shower
in the image, the major axis becomes rather poorly defined and the
angular resolution will suffer. 

\begin{figure}[hb]
  \includegraphics[width = 0.45\textwidth]{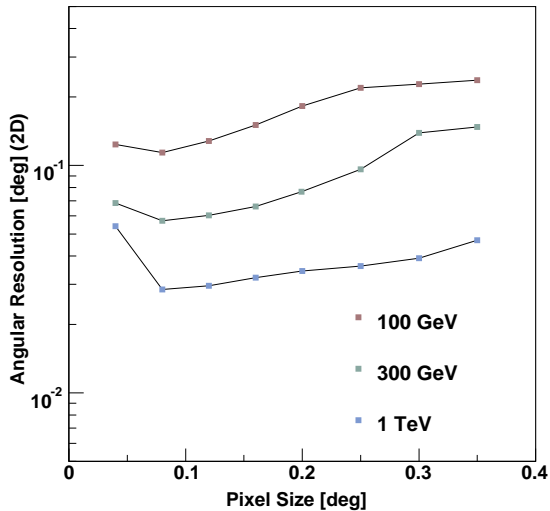}
  \caption{Angular resolution as a function of angular pixel size for
    a fixed number of pixels (36x36=1296) for the reference array.
    Since the number of pixels is kept constant, the field of view
    increases as the pixel size increases. Only in the lowest point
    for the 1 TeV shower there is actually some effect visible from
    the reduction in the field of view (from the comparison to
    Fig~\ref{fig::Fig5}). For the smallest pixel size
    (0.04$^{\circ}$), the FoV is 1.44$^{\circ}$.} \label{fig::Fig4}
\end{figure}

\begin{figure}[ht]
  \includegraphics[width = 0.45\textwidth]{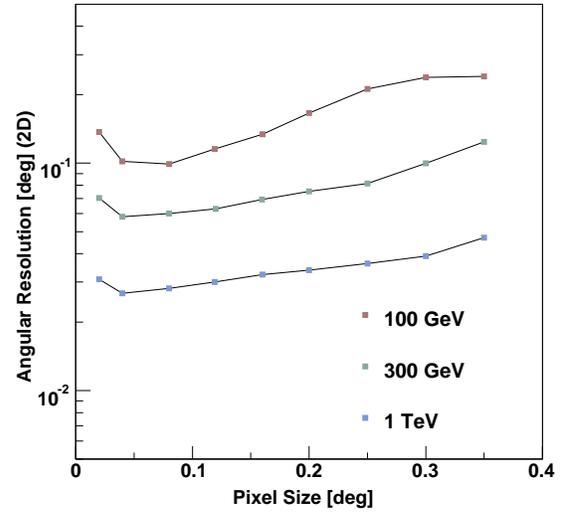}
  \caption{Angular resolution as a function of angular pixel size for
    a fixed field of view (6$^{\circ}$) for the reference array. As
    suggested already in Figure 4, the improvement with decreasing
    pixel size is rather modest, so the optimal pixel size is in the
    range of 0.05$^{\circ}$. It should be noted, that this is true for
    the Hillas analysis and does not preclude a significant
    improvement with smaller pixels and a different reconstruction
    scheme.} \label{fig::Fig5}
\end{figure}

Naturally the angular size of the individual pixels, together with
the total number of pixels determines the field of view. Each pixel
adds to the cost, and therefore we have taken the approach of fixing
the number of pixels in the camera (and therefore the total costs)
while changing the pixel size. The number of pixels was fixed at 1296
per camera.  The effect of fixing the pixel size is that while the
pixel size gets bigger, the FoV of the camera also
increases. Fig~\ref{fig::Fig4} shows the angular resolution as a
function of the pixel size for the reference array of 49
telescopes. As can be seen from this figure, the angular resolution
gets only slightly worse as the pixel size increases and the
dependence seems to be rather modest. To check whether this effect is
due to the shrinking of the field of view with decreasing pixel size,
we also kept the field of view at a constant value of 6$^{\circ}$ and
determined again the angular resolution as a function of the pixel
size. The results are shown in Fig~\ref{fig::Fig5}. As can be seen,
the behaviour is very similar to that shown in Fig~\ref{fig::Fig4},
suggesting that the increase in the field of view has a rather small
effect on the angular resolution (this has been verified independently
by varying the field of view with fixed pixel size).  Summarising
these findings, for a Hillas-type analysis, there seems to be only a
slight improvement in angular resolution when going to smaller
pixels. The optimal resolution is achieved for angular pixel sizes ~
0.05$^{\circ}$. It should be noted that at this stage, we do not see a
clear improvement in angular resolution when using more advanced
reconstruction methods than a simple Hillas-style analysis. It should
however be noted, that this is work in progress and there could be
some improvement in particular with smaller pixels when using
different techniques such as a Maximum Likelihood fit of all the
pixels in the image simultaneously.

\section{Summary}
We have devised a simple Monte-Carlo simulation scheme based on
Corsika-simulated gamma-ray showers to explore the phase-space of how
to build a future ground-based Cherenkov telescope with optimal
angular resolution.  Preliminary results point to only a modest
improvement in angular resolution with smaller angular pixel size and
no significant improvement (for 1 TeV showers from zenith) when having
more than 10 telescopes detecting the shower. We have verified this
approach by simulating a H.E.S.S.-like instrument and getting matching
angular resolution as a function of energy. The next steps are an
exploration of the phase space to determine parameters of the array to
optimise the angular resolution.

\begin{theacknowledgments}
  The authors would like to thank the members of the H.E.S.S.\, CTA
  and AGIS collaborations for help and interesting discussions. In
  particular we would like to acknowledge the help of K. Bernl{\"o}hr
  in all issues related to the MC production.
\end{theacknowledgments}



\bibliographystyle{aipproc}   % if natbib is available

\bibliography{Gamma2008_AngularResolution}

\IfFileExists{\jobname.bbl}{}
 {\typeout{}
  \typeout{******************************************}
  \typeout{** Please run "bibtex \jobname" to optain}
  \typeout{** the bibliography and then re-run LaTeX}
  \typeout{** twice to fix the references!}
  \typeout{******************************************}
  \typeout{}
 }

\end{document}